# Effect of anharmonicity on the thermal conductivity of amorphous silica

Xueyan Zhu [1,2,*], Cheng Shao [3]

[1] CAEP Software Center for High Performance Numerical Simulation, Beijing 100088, P.R. China

[2] Institute of Applied Physics and Computational Mathematics, Beijing 100088, P.R. China

[3] Department of Mechanical Engineering, The University of Tokyo, 7-3-1 Hongo, Bunkyo, Tokyo 113-8656, Japan

**Abstract**

Proper consideration of anharmonicity is important for the calculation of the thermal conductivity. However, how the anharmonicity influences the thermal conduction in amorphous materials is still an open question. In this work, we uncover the role of anharmonicity on the thermal conductivity of amorphous silica (a-$SiO_2$) by comparing the thermal conductivity predicted from the harmonic theory and the anharmonic theory. Moreover, we explore the effect of anharmonicity-induced frequency shift on the prediction of the thermal conductivity. It is found that the thermal conductivity calculated by the recently developed anharmonic theory (quasi-harmonic Green-Kubo approximation, QHGK) is higher than that by the harmonic theory developed by Allen and Feldman. The use of anharmonic vibrational frequencies also leads to a higher thermal conductivity compared with that calculated using harmonic vibrational frequencies. The anharmonicity induced frequency shifts is a mechanism for the positive temperature dependence of the thermal conductivity of a-$SiO_2$ at higher temperatures. Further investigation on mode diffusivity suggests that although anharmonicity has larger influence on locons than diffusons, the increase of the thermal conductivity due to the anharmonicity is mainly contributed by the anharmonicity induced increase of the diffusivity of diffusons. Finally, it is found that the cross-correlations between diffusons and diffusons contribute most to the thermal conductivity of a-$SiO_2$, and the locons contribute to the thermal conductivity mainly through collaboration with diffusons. These results offer new insights into the nature of the thermal conduction in a-$SiO_2$.

[*] zhu_xueyan@iapcm.ac.cn

# I. INTRODUCTION

The role of anharmonicity on the thermal conduction in amorphous solids is quite different from that in their crystalline counterparts. For crystalline materials, it is known that anharmonicity leads to the reduction of phonon mean free path, which causes the decrease of the thermal conductivities at higher temperatures [1]. However, for amorphous materials, the thermal conductivities increase monotonically with the increase in temperature [2]. The mechanism for this phenomenon and the effect of anharmonicity on the thermal conduction in amorphous materials is still an open question.

Early studies of Alexander *et al.* [3,4] and Jagannathan *et al.* [5] suggested that anharmonic interactions in amorphous materials can induce phonon-assisted hopping of localized fraction, thus leading to the increase of the thermal conductivity. Allen and Feldman [6,7] derived a thermal conductivity model (AF theory) for amorphous materials from the Kubo formula under the harmonic approximation. This harmonic theory successfully predicts the positive temperature dependence of the thermal conductivity in amorphous silicon, which is totally attributed to the increase of the heat capacity with temperature. However, later studies by Shenogin *et al.* [8] showed that positive temperature dependence of the thermal conductivity in amorphous silica (a-$SiO_2$) and polystyrene can be predicted by molecular dynamics simulations that ignore the temperature dependence of the specific heat. Their results suggest that anharmonicity is a mechanism for the rise of the thermal conductivity of amorphous materials with complex composition at higher temperatures. Although current research found a positive role of anharmonicity on the heat conduction in amorphous materials, the underlying mechanism for the influence of anharmonicity is still unclear.

Recently, several thermal conductivity models that can consider both anharmonicity and disorder have been proposed, which provide tools for investigating the effect of anharmonicity on the thermal transport in amorphous materials. Lv and Henry [9] developed the Green-Kubo Modal Anaysis (GKMA) method, which combines Green-Kubo formula with the lattice dynamics formulism. This method uses molecular dynamics (MD) simulations to obtain the modal contributions to the heat flux, thus naturally includes anharmonicity and disorder. The ability to consider modal contributions allows it to apply a quantum correction to the heat capacity. GKMA has been used to explore the effect of anharmonicity on the thermal conduction in a-SiO$_2$ [10]. It was found that locons are largely responsible for the rise in the thermal conductivity above room temperature, and may contribute to the thermal conductivity through collaboration with other modes due to anharmonicity. Simoncelli *et al.* [11] derived an equation that accounts for both anharmonicity and disorder from the Wigner phase space formulation of quantum mechanics. This equation can reduce to the expression derived by Peierls-Boltzmann transport equation for anharmonic crystals and the AF theory for harmonic amorphous materials. Isaeva *et al.* [12] obtained an equation that can also consider both anharmonicity and disorder through combing the Green-Kubo (GK) theory and a quasi-harmonic description of lattice vibrations. This equation was dubbed as the quasi-harmonic Green-Kubo approximation (QHGK) for the thermal conductivity.

In this work, QHGK is chosen to compare with AF theory for exploring the effect of anharmonicity on the thermal conductivity of a-SiO$_2$. This is because the approaches for deriving QHGK and AF theory are similar. In addition, the thermal conductivities computed using harmonic vibrational frequencies are compared with those using anharmonic vibrational frequencies. The mechanism for the rise of the thermal conductivity at higher temperatures is revealed through these comparisons. Moreover, the mode diffusivities with and without considering anharmonicity are calculated and compared for uncovering the effect of anharmonicity on diffusons and locons, respectively.

The rest of this manuscript is organized as follows. The theoretical formulations used in this work are reviewed in Sec. II. The computational details are presented in Sec. III. The calculated vibrational mode properties are presented in Sec. IV, including the inverse participation ratio (IPR), the phonon density of states (DOS), the vibrational frequencies and the mode lifetimes at different temperatures. In Sec. V, comparison is made between the thermal conductivities and diffusivities calculated using QHGK and AF theory. In Sec. VI, comparison is made between the thermal conductivities and diffusivities calculated using anharmonic frequencies and harmonic frequencies. The conclusions are presented in Sec. VII.

## II. THEORETICAL FORMULATIONS

Two models are used in this work to uncover the role of anharmonicity on the thermal conduction in a-SiO$_2$, including AF theory [6] and QHGK [12]. AF theory was derived based on Kubo formula using harmonic heat current operator. QHGK was derived based on Green-

Kubo formula starting from the harmonic heat flux, and then anharmonicity was introduced through the linewidths of the vibrational normal modes when calculating the heat-flux auto-correlation function.

The formula of AF theory [6] is expressed by

$$k = \frac{1}{V}\sum_i C_i(T) D_i. \tag{1}$$

In the above equation, $V$ is the volume of the system. $C_i(T)$ is the specific heat of mode $i$,

$$C_i = k_B \frac{[\hbar\omega_i/(k_B T)]^2 e^{\hbar\omega_i/(k_B T)}}{[e^{\hbar\omega_i/(k_B T)} - 1]^2}, \tag{2}$$

where $\omega_i$ is the vibrational frequency, $k_B$ is the Boltzmann constant, and $T$ is the temperature. $D_i$ is the the diffusivity of mode $i$,

$$D_i = \frac{\pi V^2}{3\hbar^2 \omega_i^2} \sum_j^{\neq i} |S_{ij}|^2 \delta(\omega_i - \omega_j). \tag{3}$$

where, $\delta$ is the Dirac delta function, which is approximated by Lorentzian function,

$$\delta(\omega_i - \omega_j) = \frac{\eta}{\pi[(\omega_i - \omega_j)^2 + \eta^2]}, \tag{4}$$

with $\eta$ the line broadening parameter. $\mathbf{S}_{ij}$ is the heat current operator:

$$\mathbf{S}_{ij} = \frac{\hbar}{2V} \mathbf{v}_{ij} (\omega_i + \omega_j). \tag{5}$$

The velocity operator $\mathbf{v}_{ij}$ is:

$$\mathbf{v}_{ij} = \frac{i}{2\sqrt{\omega_i \omega_j}} \sum_{\alpha,\beta} \sum_{m,\kappa,\kappa'} \frac{\Phi_{\kappa'\kappa}^{\beta\alpha}(0,m)}{\sqrt{m_\kappa m_{\kappa'}}} e_\kappa^{i,\alpha} e_{\kappa'}^{j,\beta} (\mathbf{R}_m + \mathbf{R}_{\kappa\kappa'}), \tag{6}$$

where, $\Phi_{\kappa'\kappa}^{\beta\alpha}(0,m)$ is the non-Hermitian force constants, $e_\kappa^{i,\alpha}$ is the phonon eigenvector, $m_\kappa$ is the mass of atom $\kappa$, $\mathbf{R}_m$ is the position of cell $m$, and $\mathbf{R}_{\kappa\kappa'}$ is the distance between atom $\kappa$ and atom $\kappa'$ in a cell.

QHGK approximation [12] for the thermal conductivity is

$$k^{\alpha\beta} = \frac{1}{V} \sum_{ij} c_{ij} v_{ij}^\alpha v_{ij}^\beta \tau_{ij}, \tag{7}$$

$$c_{ij} = \frac{\hbar \omega_i \omega_j}{T} \frac{f_{0i} - f_{0j}}{\omega_j - \omega_i}, \tag{8}$$

$$\tau_{ij} = \frac{\Gamma_i + \Gamma_j}{\left(\Gamma_i + \Gamma_j\right)^2 + \left(\omega_i - \omega_j\right)^2} + O\left(\varepsilon^2\right). \tag{9}$$

$f_{0i}$ is the Bose-Einstein occupation number of the $i$-th normal mode. $\Gamma_i$ is the linewidth of mode $i$, which is related to the anharmonic lifetime by

$$\tau_i = \frac{1}{2\Gamma_i}. \tag{10}$$

In Eq. (7), the summation of terms with $i = j$ is the auto-correlation contribution to the thermal conductivity, while that with $i \neq j$ is the cross-correlation contribution.

To validate the results of QHGK, the thermal conductivities computed by QHGK in the limit of classical specific heat are compared with that calculated by Green-Kubo formula based on MD simulations. In addition, it is found that the results of QHGK is in agreement with the thermal conductivities calculated by the anharmonic theory derived by Simoncelli, Marzari, and Mauri [11].

The anharmonic vibrational frequencies and lifetimes are computed using MD simulation-based normal mode decomposition (NMD) method [13-15]. In NMD, the velocities of each atom from MD simulations are first projected onto the normal mode:

$$Q(\mathbf{k}, \nu, t) = \sum_\alpha^3 \sum_j^n \sum_l^N \sqrt{\frac{m_j}{N}} u_{jl}^\alpha(t) e_j^{\alpha*}(\mathbf{k}, \nu) \exp(-i\mathbf{k} \cdot \mathbf{r}_{lj}), \tag{11}$$

where, $N$ is the total number of unit cells, $m_j$ is the mass of atom $j$, $u$ is the displacement, and $e$ is the phonon eigenvector. Then, the spectral energy density (SED) is calculated through the Fourier transform of the time derivative of the normal mode:

$$\Phi(\mathbf{k}, \nu, \omega) = \left|F\left(\dot{Q}(\mathbf{k}, \nu, t)\right)\right|^2 = \left|\int_0^{+\infty} \dot{Q}(\mathbf{k}, \nu, t) e^{-i\omega t} dt\right|^2. \tag{12}$$

The anharmonic vibrational frequencies and lifetimes are obtained through fitting the SED by a Lorentzian function:

$$\Phi(\mathbf{k}, \nu, \omega) = \frac{C(\mathbf{k}, \nu)}{\left[\omega - \omega^A(\mathbf{k}, \nu)\right]^2 + \left[\Gamma(\mathbf{k}, \nu)\right]^2}, \tag{13}$$

where, $\omega$ is the harmonic frequency, $\omega^A$ is the anharmonic frequency, and $\Gamma$ is the linewidth.

For a-SiO$_2$, the atomic structure is disordered. Therefore, the NMD is carried out only at the gamma point ($\mathbf{k} = 0$).

## III. COMPUTAIONAL DETAILS

MD simulations implemented in LAMMPS [16] are performed for computing the vibrational properties and thermal conductivities of a-SiO$_2$. The van Beest-Krammer-van Santen (BKS) potential modified by adding a 24-6 Lennard-Jones (LJ) potential [17-19] is employed. The cutoffs of Buckingham and LJ potentials are set to 10 Å and 8.5 Å, respectively. The electrostatic interactions are calculated through the Wolf summation method with a cutoff of 12 Å and a damping factor of 0.223 Å$^{-1}$.

All the simulations are carried out with periodic boundary conditions and a timestep of 0.905 fs. The melt-quench procedure is used to obtain the atomic configurations of a-SiO$_2$ at different temperatures. First, a system containing 648 atoms for a-SiO$_2$ is melted to 8000 K in NPT ensemble. Then, the system is cooled at a rate of 2.21 ~ 3.37 K/ps to 1900 K, 1500 K, 1300 K, 1100 K, 900 K, 700 K, 500 K, 300 K, 200 K, and 100 K, respectively. When cooled to 1100 K, a-SiO$_2$ is annealed in NPT ensemble for 9 ns to remove metastability before continuing the cooling. After the cooling process, the system is equilibrated for 1.8 ns in NPT ensemble, 0.9 ns in NVT ensemble, and 0.9 ns in NVE ensemble.

For calculating the thermal conductivities through Green-Kubo formula, the timestep is set to 0.2 fs, and the MD simulations are carried out for 8 ns in NVE ensemble. The autocorrelation function of the heat flux is calculated every 2 fs. The thermal conductivity is obtained by integrating the autocorrelation function of the heat flux.

To perform the NMD, the equilibrium configuration of a-SiO$_2$ is first obtained through an energy minimization. Then, the system is equilibrated with a time step of 0.2 fs in NVT ensemble for 0.2 ns at the targeted temperature, and additional 0.2 ns in NVE ensemble. After the equilibrium process, the velocities of each atom are dumped every 5 fs in NVE ensemble for a time span of 1 ns for postprocessing.

Force constants are calculated by GULP [20] based on the equilibrium configurations of a-SiO$_2$. Then, the harmonic vibrational properties are obtained through lattice dynamics calculations using the Phonopy code [21].

## IV. VIBRATIONAL MODE PROPERTIES

The IPR and harmonic DOS are computed as shown in Fig. 1. The equation for calculating the IPR is

$$\frac{1}{p(\mathbf{k},\nu)} = \sum_j \left[ \sum_\alpha \left( e_j^\alpha (\mathbf{k},\nu) \right)^2 \right]^2, \tag{14}$$

where $e_j^\alpha$ is the eigenvector. The IPR characterizes the extent of localization for a specific mode. For the vibration localized on a single atom, $1/p$ would be 1, while $1/p$ would be $1/N$ if the vibrations were equally distributed on all atoms. Fig. 1(a) indicates that locons are located in two frequency regions, $175 - 206$ rads/ps and above 234 rads/ps, which are shaded by light red. The other modes are delocalized, including propagons and diffusons. Larkin and McGaughey [22] have reported the propagons/diffusons cutoff frequency of 4.55 rad/ps. Here we use the same cutoff frequency. The region of propagons are shaded by light blue in Fig. 1, and the region of diffusons are not shaded. These results indicate that the vibrational modes of a-SiO$_2$ are dominated by diffusons. Due to the limited size of our simulated system, the lowest frequency predicted in this work is 7.31 rads/ps. Therefore, only diffusons and locons is considered in this work, which is reasonable because the contribution of propagons to the total thermal conductivity of a-SiO$_2$ is negligible according to the previous investigations [22].

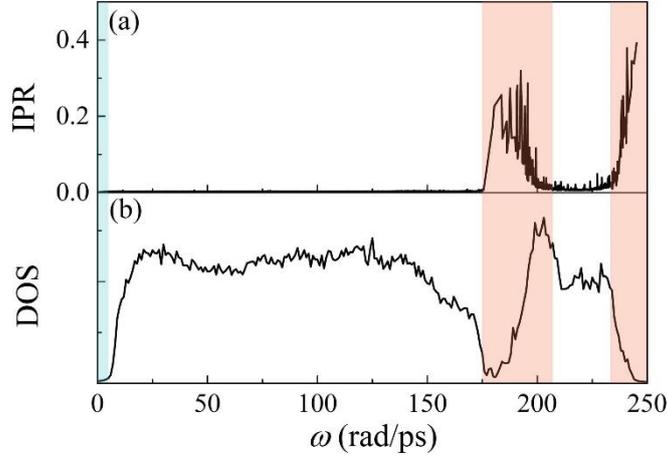

FIG. 1. The inverse participation ratio (IPR) and density of states (DOS) of vibrational modes (300 K) in a-SiO$_2$.

The DOS is calculated through:

$$\mathrm{DOS}(\omega) = \sum_i \delta(\omega_i - \omega), \tag{15}$$

and plotted in Fig. 1(b). For diffusons, the DOS is constant over most of the range of frequencies. The DOS of locons increases with frequency in the range of $175 - 206$ rads/ps, while decreases in the range of $234 - 245$ rads/ps. Locons only constitute approximately 10% of the total vibrational modes.

The method of NMD described in Section II is used for calculating the anharmonic vibrational frequencies and lifetimes at different temperatures. Using the anharmonic vibrational frequencies, the density of states of different temperatures are computed. As shown in Fig. 2, the anharmonic vibrational frequencies decrease with the increase in temperature, and the frequency shifts are more pronounced for locons. Lv *et al.* [10] also reported the lowering of vibrational frequencies of a-SiO$_2$ at higher temperatures.

The vibrational mode lifetimes are plotted in Fig. 3. The lifetimes of diffusons slowly decrease with frequency, and then increase near diffusons/locons cutoff (175 rad/ps) and a peak forms. With the increase in temperature, the lifetimes decrease.

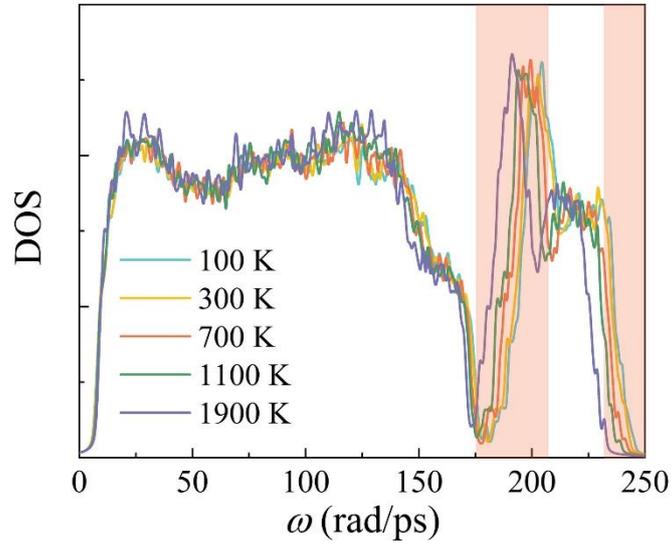

FIG. 2. The DOS of anharmonic vibrational modes at different temperatures.

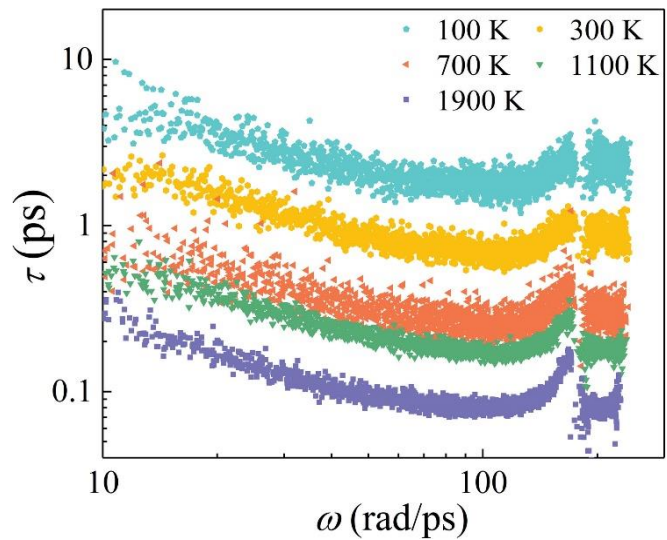

FIG. 3. The vibrational mode lifetimes of a-SiO$_2$ at different temperatures.

**V. COMPARISION BETWEEN AF THEORY AND QHGK**

To investigate the anharmonic effect on the thermal conductivities of a-SiO$_2$, the analysis can be carried out in two levels. First, one can compare the results from AF theory and QHGK. AF theory was derived based on harmonic approximation, which can only be used for predicting the thermal conductivities of stiff material at low temperatures [6]. QHGK is applicable for systems with strong anharmonicity at high temperatures due to the introduction of vibrational lifetimes [12]. Second, since the vibrational frequencies change with temperature due to anharmonicity, using the temperature dependent frequencies allows us to consider the anharmonic effect more accurately. In this section, we compare the results of AF theory and QHGK. In the next section, the thermal conductivities calculated using temperature dependent frequencies will be compared with that calculated using harmonic frequencies.

The result of AF theory (Eqs. (1-6)) is sensitive to the choice of the line broadening parameter $\eta$. The effect of $\eta$ on the thermal conductivity is shown in Fig. 4. With the increase in $\eta$, the thermal conductivity first increases to a peak value at $\eta = 0.2$ meV, then decreases. The maximum thermal conductivities predicted by AF theory are lower than that calculated by QHGK at all temperatures. The difference between the results of QHGK and AF theory becomes larger at higher temperatures. This result indicates that more careful consideration of anharmonicity in the model leads to the increase in the thermal conductivities, especially at high temperatures.

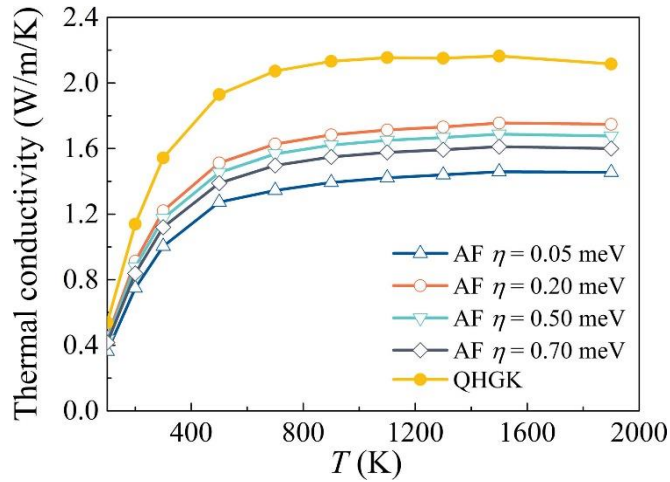

FIG. 4. The thermal conductivities calculated by AF theory and QHGK using harmonic vibrational frequencies. $\eta$ is the broadening parameter of AF theory.

The classical AF theory and QHGK can both be written in the form of Eq. (1) with $C_i = k_B$. Therefore, the difference between the thermal conductivities calculated by these two models lies in the difference between the diffusivities. The mode diffusivities of AF theory and QHGK can be expressed by

$$D_i^{AF} = \frac{1}{3}\sum_{j}^{\neq i} D_{ij}^{AF} = \frac{1}{3}\sum_{j}^{\neq i} |v_{ij}|^2 \left[\frac{\pi}{4}\left(\frac{\omega_j}{\omega_i}+1\right)^2\right]\delta(\omega_j - \omega_i) \quad (16)$$

and

$$D_i^{QHGK} = \frac{1}{3}\sum_{j} D_{ij}^{QHGK} = \frac{1}{3}\sum_{j} |v_{ij}|^2 \tau_{ij}, \quad (17)$$

respecitvely.

The comparison between the mode diffusivties of AF theory and QHGK is shown in Fig. 5(a). The diffusivites computed by QHGK are generally larger than that by AF theory. The absolute and relative difference of the diffusivities between these two models are plotted in Fig. 5(b). The absolute diffusivity difference of diffusons (4.55 – 175 rads/ps and 206 – 234 rads/ps) is generally larger than that of locons (175 – 206 rads/ps and above 234 rads/ps). However, the relative diffusivity difference of locons is much higher than that of diffusons. The relative contributions of diffusons and locons to the total diffusivity difference are calculated by

$$\Delta D_{\text{diffusons}}^{\text{QHGK-AF}} = \frac{\sum_i^{\text{diffusons}}\left(D_i^{QHGK} - D_i^{AF}\right)}{\sum_i^{\text{total}}\left(D_i^{QHGK} - D_i^{AF}\right)} \quad (18)$$

and

$$\Delta D_{\text{locons}}^{\text{QHGK-AF}} = \frac{\sum_i^{\text{locons}}\left(D_i^{QHGK} - D_i^{AF}\right)}{\sum_i^{\text{total}}\left(D_i^{QHGK} - D_i^{AF}\right)}, \quad (19)$$

respectively. The variatons of $\Delta D_{\text{diffusons}}^{\text{QHGK-AF}}$ and $\Delta D_{\text{locons}}^{\text{QHGK-AF}}$ with temperature are plotted in Fig. 6. The relative contributions of diffusons to the total diffusivity difference $\Delta D_{\text{diffusons}}^{\text{QHGK-AF}}$ accounts for 94% – 99% of the total diffusivity difference, while the relative contributions of locons $\Delta D_{\text{locons}}^{\text{QHGK-AF}}$ only accounts for 1.0% – 5.5% of the total diffusivity difference. With the increase in temperature, $\Delta D_{\text{diffusons}}^{\text{QHGK-AF}}$ decreases, while $\Delta D_{\text{locons}}^{\text{QHGK-AF}}$ increases. These results indicate that although the anharmonicity has larger influence on locons than diffusons, the increase of the diffusivites and the thermal conductivites caused by the anharmonicity are mostly contributed by the anharmonic diffusons. This is because the diffusons account for 90% of the total vibrational modes, and the diffusivities of diffusons are much larger than that of locons.

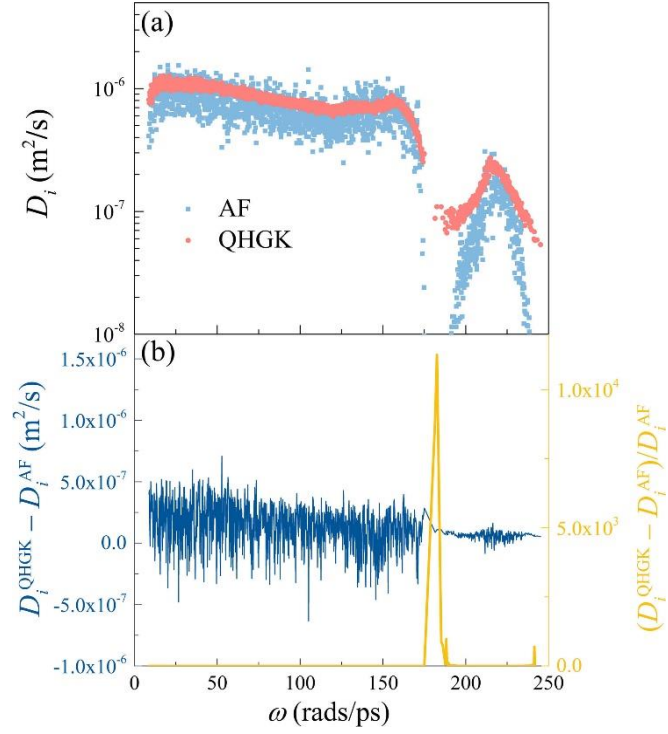

FIG. 5. (a) The diffusivities of vibrational modes in a-SiO$_2$ at 1900 K calculated by AF theory (Eq. (16)) and QHGK (Eq. (17)), respectively. (b) The absolute (left) and relative difference (right) between diffusitives calcualted by AF theory and QHGK.

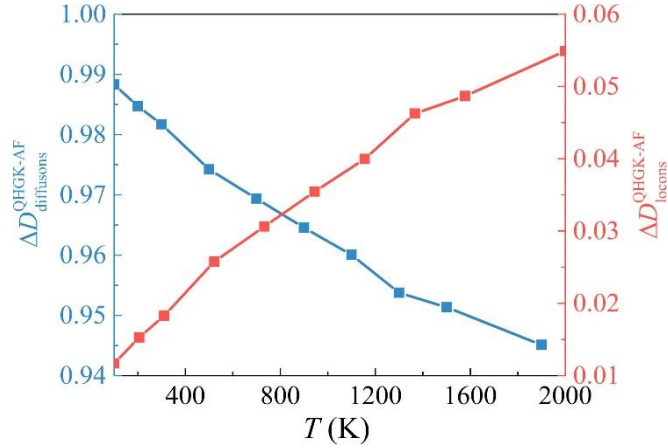

FIG. 6. The relative contributions of diffusons (left) and locons (right) to the total diffusivity difference between AF theory and QHGK.

## VI. EFFECT OF ANHARMONIC VIBRATIONAL FREQUENCIES ON THE THERMAL CONDUCTIVITY

Due to anharmonicity, the vibrational frequencies decrease with the increase in temperature as shown in Fig. 2. Using the temperature dependent frequencies (anharmonic frequencies) allows us to consider the anharmonic effect more accurately for the calculation of the thermal conductivity. Comparison is made between the thermal conductivities calculated using harmonic and anharmonic frequencies. The results are plotted in Fig. 7. The blue line shows the thermal conductivities computed by QHGK using harmonic frequencies and classical specific heat, while the yellow line is the thermal conductivities computed by QHGK using anharmonic frequencies and classical specific heat. The thermal conductivities calculated using anharmonic frequencies are higher than that calculated using harmonic frequencies at all temperatures, and the difference increases with the increase in temperature. This result indicates that the anharmonicity induced frequency shifts have positive contribution to the thermal conduction in a-SiO$_2$.

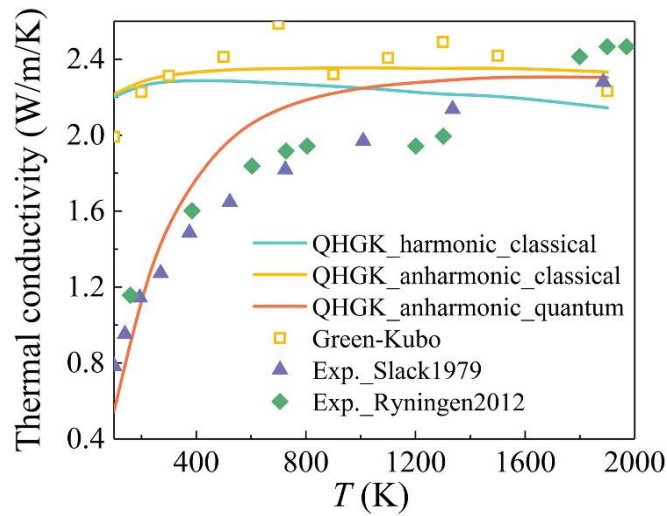

FIG. 7. The thermal conductivities calculated by QHGK using harmonic or anharmonic vibrational frequencies and classical or quantum specific heat, and their comparisons with the results of Green-Kubo formula based on MD simulations and experimental data [23,24].

The mode diffusivities computed using harmonic and anharmonic frequencies are compared and the results are shown in Fig. 8(a). Anharmonic diffusivity is larger than harmonic diffusivity for all vibrational modes. Fig. 8(b) shows the absolute and relative difference between the anharmonic and harmonic diffusivities as a function of frequency. The absolute diffusivity difference of diffusons is generally larger than that of locons, while the relative diffusivity difference of locons is larger than that of diffusons. The relative contributions of

diffusons and locons to the total diffusivity difference are calculated by

$$\Delta D_{\text{diffusons}}^{\text{anharmo-harmo}} = \frac{\sum_{i}^{\text{diffusons}} \left( D_i^{\text{anharmo}} - D_i^{\text{harmo}} \right)}{\sum_{i}^{\text{total}} \left( D_i^{\text{anharmo}} - D_i^{\text{harmo}} \right)} \tag{20}$$

and

$$\Delta D_{\text{locons}}^{\text{anharmo-harmo}} = \frac{\sum_{i}^{\text{locons}} \left( D_i^{\text{anharmo}} - D_i^{\text{harmo}} \right)}{\sum_{i}^{\text{total}} \left( D_i^{\text{anharmo}} - D_i^{\text{harmo}} \right)}, \tag{21}$$

respectively. In the above equations, $D_i^{\text{anharmo}}$ is the diffusivity calculated by Eq. (17) using anharmonic frequencies, while $D_i^{\text{harmo}}$ that calculated using harmonic frequencies. The variatons of $\Delta D_{\text{diffusons}}^{\text{anharmo-harmo}}$ and $\Delta D_{\text{locons}}^{\text{anharmo-harmo}}$ with temperature are shown in Fig. 9. $\Delta D_{\text{diffusons}}^{\text{anharmo-harmo}}$ accounts for 97% − 99% of the total diffusivity difference, while $\Delta D_{\text{locons}}^{\text{anharmo-harmo}}$ only accounts for 1.0% − 3.0%. With the increase in temperature, $\Delta D_{\text{diffusons}}^{\text{anharmo-harmo}}$ decreases, while $\Delta D_{\text{locons}}^{\text{anharmo-harmo}}$ increases. These results indicate that the increase of the diffusivites and thermal conductivites induced by anharmonic frequencies is mostly contributed by the anharmonic diffusons, while the anharmonicity has larger influence on locons than diffusons.

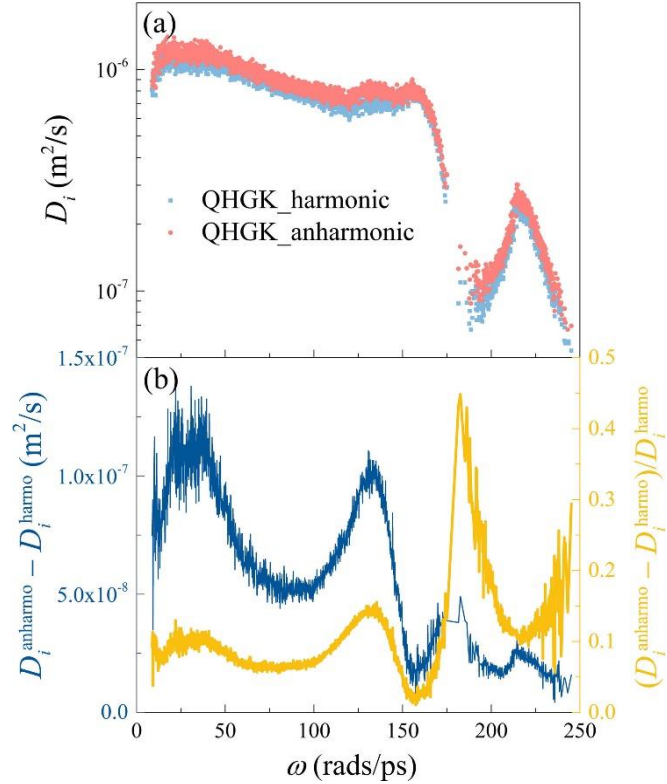

FIG. 8. (a) The diffusivities of modes in a-SiO$_2$ at 1900 K calculated by QHGK (Eq. (17)) using harmonic and anharmonic vibrational frequencies, respectively. (b) The absolute (left) and relative difference (right) between diffusitives calcualted using anharmonic and harmonic vibrational frequencies.

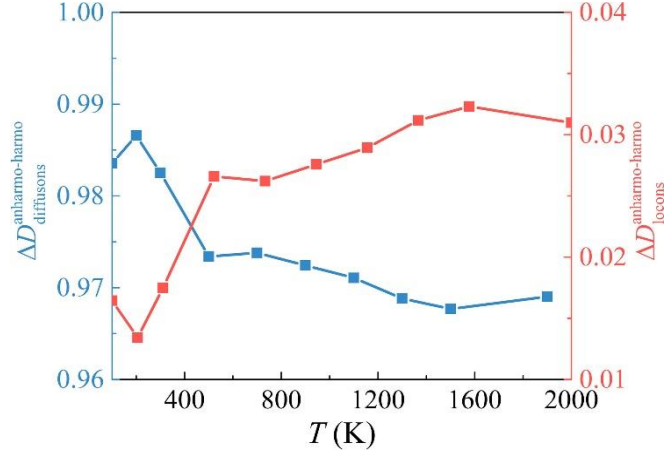

FIG. 9. The relative contributions of diffusons (left) and locons (right) to the total difference between the diffusivities calculated by anharmonic and harmonic vibrational frrequencies.

The comparisons between the thermal conductivities calculated by different methods shown in Fig. 7 reveal the mechanisms for the positive temperature dependence of the thermal conductivity of a-SiO$_2$. If the harmonic frequencies and classical specific heat were used, the thermal conductivity decreases at high temperatures (blue line in Fig. 7). The use of temperature dependent vibrational frequencies leads to a slight increase of the thermal conductivity with temperature (yellow line in Fig. 7), which is in good agreement with the thermal conductivities computed by Green-Kubo formula based on MD simulations (yellow hollow points). This indicates that the anharmonicity induced temperature dependence of the vibrational frequencies is a mechanism for the positive temperature dependence of the thermal conductivity. Furthermore, the addition of quantum specific heat leads to a steeper temperature dependence of the thermal conductivity (red line in Fig. 7), which is in reasonable agreement with experimental data.

To further investigate the mechanism for the heat transfer in a-SiO$_2$, the contributions of auto-correlation and cross-correlation of diffusons and locons to the thermal conductivity are calculated. It is found that the auto-correlations of both diffusons and locons are zero. The contributions of cross-correlations of diffusons-diffusons, diffusons-locons, and locons-locons to the total thermal conductivities are plotted in Fig. 10. The cross-correlations between diffusons and diffusons, and between diffusons and locons, contribute 97.37% ~ 99.99% and 2.22% ~ 0.01% to the total thermal conductivity, respectively. The cross-correlations between locons and locons are negligible. These results indicate that heat is transferred in a-$SiO_2$ mainly through the interaction between diffusons and diffusons. Locons contribute to the heat transfer mainly through collaboration with diffusons.

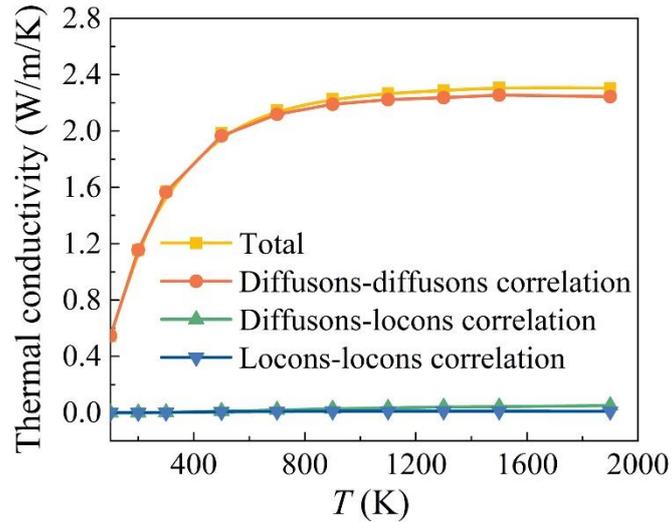

FIG. 10. The contributions of cross-correlations between diffusons and diffusons, between diffusons and locons, and between locons and locons to the total thermal conductivity, respectively.

The effects of anharmonicity on the thermal conductivity of a-$SiO_2$ are summarized using colored bars in Fig. 11. Anharmonicity has positive contribution to the thermal conduction in a-$SiO_2$. Consideration of anharmonicity in the thermal conductivity model (QHGK) enhances the contribution of cross-correlations of diffusons-locons to the thermal conductivity. Diffusons-diffusons cross-correlations dominates the heat conduction in a-$SiO_2$.

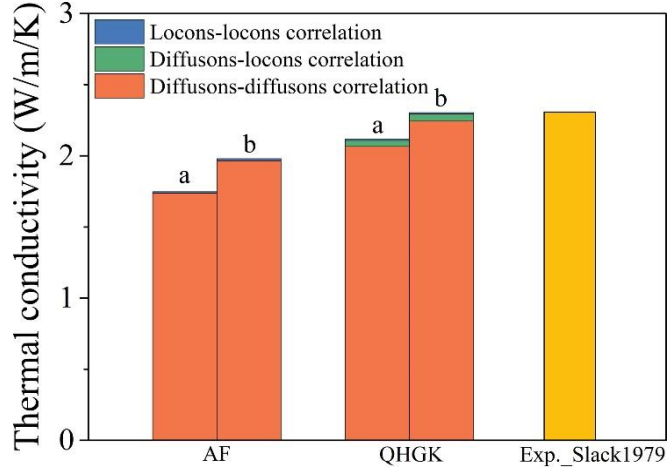

FIG. 11. Thermal conductivity of a-SiO$_2$ at 1900 K computed by AF theory and QHGK, respectively, and a comparison with experimental data [23]. The symbol (a) represents the results calculated using harmonic vibrational frequencies, while (b) represents the results calculated using anharmonic vibrational frequencies.

## VII. CONCLUSION

We investigated the effect of anharmonicity on the thermal conduction in a-SiO$_2$ from two perspectives. First, comparison is made between QHGK and AF theory. QHGK is a thermal conductivity model that can consider both anharmonicity and disorder. AF theory is also applicable for amorphous materials, but it ignores the effects of anharmonicity in the atomic interactions. It is found that the thermal conductivities calculated by QHGK are larger than the predictions of AF theory. Second, comparison is made between thermal conductivities calculated using anharmonic and harmonic vibrational frequencies. It is found that the use of anharmonic vibrational frequencies results in a higher thermal conductivity compared with that calculated using harmonic vibrational frequencies. The temperature dependence of the anharmonic vibrational frequencies is a mechanism for the positive temperature dependence of the thermal conductivity. Further investigation of mode diffusivities indicates that the enhancement of the thermal conductivity due to the anharmonicity is mainly contributed by the anharmonic diffusons, though the anharmonicity has larger influence on locons than diffusons. This is because diffusons have larger mode diffusivities than locons, and diffusons constitute 90% of the total vibrational modes. Finally, it is found that cross-correlation between diffusons and diffusons is the major mechanism for the heat conduction in a-SiO$_2$, and locons transfer the heat mainly through collaboration with diffusons.

## ACKNOWLEDGMENTS

The authors gratefully thank Hua Bao for many insightful suggestions and reviews of this manuscript. This work was supported by National Natural Science Foundation of China (Grant No. 12005019).

**References**

[1] V. Murashov and M. A. White, in *Thermal conductivity: theory, properties, and applications*, edited by T. M. Tritt (Kluwer Academic / Plenum Publishers, New York, 2004), pp. 93.

[2] A. I. Krivchikov and A. Jeżowski, Thermal conductivity of glasses and disordered crystals, arXiv:2011.14728 (2020).

[3] S. Alexander, C. Laermans, R. Orbach, and H. M. Rosenberg, Fracton interpretation of vibrational properties of cross-linked polymers, glasses, and irradiated quartz, Physical Review B **28**, 4615 (1983).

[4] S. Alexander, O. Entin-Wohlman, and R. Orbach, Phonon-fracton anharmonic interactions: The thermal conductivity of amorphous materials, Physical Review B **34**, 2726 (1986).

[5] A. Jagannathan, R. Orbach, and O. Entin-Wohlman, Thermal conductivity of amorphous materials above the plateau, Physical Review B **39**, 13465 (1989).

[6] P. B. Allen and J. L. Feldman, Thermal conductivity of disordered harmonic solids, Physical Review B **48**, 12581 (1993).

[7] J. L. Feldman, M. D. Kluge, P. B. Allen, and F. Wooten, Thermal conductivity and localization in glasses: Numerical study of a model of amorphous silicon, Physical Review B **48**, 12589 (1993).

[8] S. Shenogin, A. Bodapati, P. Keblinski, and A. J. H. McGaughey, Predicting the thermal conductivity of inorganic and polymeric glasses: The role of anharmonicity, Journal of Applied Physics **105**, 034906 (2009).

[9] W. Lv and A. Henry, Direct calculation of modal contributions to thermal conductivity via Green-Kubo modal analysis, New Journal of Physics **18**, 013028 (2016).

[10] W. Lv and A. Henry, Non-negligible Contributions to Thermal Conductivity From Localized Modes in Amorphous Silicon Dioxide, Scientific Reports **6**, 35720 (2016).

[11] M. Simoncelli, N. Marzari, and F. Mauri, Unified theory of thermal transport in crystals and glasses, Nature Physics **15**, 809 (2019).

[12] L. Isaeva, G. Barbalinardo, D. Donadio, and S. Baroni, Modeling heat transport in crystals and glasses from a unified lattice-dynamical approach, Nature Communications **10**, 3853 (2019).

[13] M. T. Dove, *Introduction to Lattice Dynamics* (Cambridge University Press, 1993).

[14] A. J. H. McGaughey and J. M. Larkin, Predicting Phonon Properties from Equilibrium Molecular Dynamics Simulations, Annual Review of Heat Transfer **17**, 49 (2014).

[15] C. Shao and J. Shiomi, Negligible contribution of inter-dot coherent modes to heat conduction in

quantum-dot superlattice, Materials Today Physics **22**, 100601 (2022).

[16] S. Plimpton, Fast Parallel Algorithms for Short-Range Molecular Dynamics, Journal of Computational Physics **117**, 1 (1995).

[17] B. W. H. van Beest, G. J. Kramer, and R. A. van Santen, Force fields for silicas and aluminophosphates based on ab initio calculations, Physical Review Letters **64**, 1955 (1990).

[18] Y. Guissani and B. Guillot, A numerical investigation of the liquid–vapor coexistence curve of silica, The Journal of Chemical Physics **104**, 7633 (1996).

[19] A. J. H. McGaughey and M. Kaviany, Thermal conductivity decomposition and analysis using molecular dynamics simulations: Part II. Complex silica structures, International Journal of Heat and Mass Transfer **47**, 1799 (2004).

[20] J. D. Gale and A. L. Rohl, The General Utility Lattice Program (GULP), Molecular Simulation **29**, 291 (2003).

[21] A. Togo and I. Tanaka, First principles phonon calculations in materials science, Scripta Materialia **108**, 1 (2015).

[22] J. M. Larkin and A. J. H. McGaughey, Thermal conductivity accumulation in amorphous silica and amorphous silicon, Physical Review B **89**, 144303 (2014).

[23] G. A. Slack, in *Solid State Physics*, edited by H. Ehrenreich, F. Seitz, and D. Turnbull (Academic Press, 1979), pp. 1.

[24] B. Ryningen, M. P. Bellmann, R. Kvande, and O. Lohne, The Effect of Crucible Coating and The Temperature Field on Minority Carrier Lifetime in Directionally Solidified Multicrystalline Silicon Ingots, in *27th European Photovoltaic Solar Energy Conference and Exhibition* (2012), pp. 926.